\documentclass[aps,pra,twocolumn,showpacs,amsmath,amssymb]{revtex4-1}

\usepackage{amsthm} \usepackage{epsfig,pstricks-add}
\usepackage{latexsym,wasysym}

\newcommand{\e}{\varepsilon} \newcommand{\re}{\rightarrow}

\newtheorem{thm}{Theorem} \newtheorem{lem}{Lemma}
\newtheorem{definition}{Definition} \newtheorem{cor}{Corollary}
\newtheorem{ex}{Example}

\begin{document}

\title{Bipartite Units of Nonlocality}

\author{Manuel Forster, Stefan Wolf} \affiliation{Computer Science
  Department, ETH Z\"urich, CH-8092 Z\"urich, Switzerland}

\date{\today}

\begin{abstract}
  Imagine a task in which a group of separated players aim to simulate a
  statistic that violates a Bell inequality. Given measurement choices
  the players shall announce an output based solely on the results of
  local operations -- which they can discuss before the separation -- on
  shared random data and shared copies of a so-called {\em unit}
  correlation. In the first part of this article we show that in such a
  setting the simulation of any bipartite correlation, not containing
  the possibility of signaling, can be made arbitrarily accurate by
  increasing the number of shared Popescu-Rohrlich (PR) boxes. This
  establishes the PR box as a simple asymptotic unit of bipartite
  nonlocality. In the second part we study whether this property extends
  to the multipartite case. More generally, we ask if it is possible for
  separated players to asymptotically reproduce any nonsignaling
  statistic by local operations on bipartite unit correlations. We find
  that non-adaptive strategies are limited by a constant accuracy and
  that arbitrary strategies on $n$ resource correlations make a mistake
  with a probability greater or equal to $c/n$, for some constant $c$.
\end{abstract}

\maketitle

\section{Introduction}

The correlation in the outputs of certain quantum experiments on pairs
of separated particles cannot be explained by information shared before
the separation. This property is called {\em quantum nonlocality} and
manifests itself in the violation of Bell
inequalities~\cite{Bell-1964,CHSH-1969}. Nonlocal quantum correlations
have opened new opportunities for information processing such as, for
example, quantum key
distribution~\cite{Hardy05,PhysRevLett.98.230501,PhysRevLett.102.140501,
  hanggi-2010} with device-independent secrecy. Correlations that do not
offer the possibility of signaling can be represented, when measurement
and outcome dimensions are fixed, by so-called nonsignaling polytopes of
which quantum correlations are a proper
subset~\cite{cirelson-1980}. Some elements in these convex sets are more
useful for distributed tasks than
others~\cite{vandam-2005,Linden-2006,brassard-2006,brunner2009}. Thus, a
number of theoretical questions on the relationship and possible
reductions between different nonsignaling correlations have emerged. In
the context of quantum correlations the singlet state has been
established as a unit of entanglement: A supply of singlets can be
transformed into any other bipartite pure state by local operations and
classical communication, and vice versa~\cite{bennett-1996}. This
reversibility partly relies on asymptotic transformations and does not
hold in general, that is, for multipartite states or bipartite mixed
states. Summing up, we have that any entangled state can be approximated
from sufficiently many copies of the singlet state.

An analogous question when the objects are not quantum states, but
general nonlocal correlations, motivates the search for a {\em unit of
  nonlocality}. More specifically, the identification of a nontrivial
set of correlations from which any other nonsignaling statistic can be
derived, is
intended~\cite{barret-2005,jones-2005,pironio-2005,dupuis-2007}. The
following game illustrates the problem: In the initial phase a group of
players is given a description of certain input-output
correlations. They are allowed to discuss a strategy but are then
separated in order to prevent them from communicating. Now the test
phase begins. Each player is given a secret input and announces an
output. This is repeated many times independently. The objective of the
players is to minimize the distance of their input-output distribution
from the described target correlations.

Any {\em local} correlation can be simulated perfectly by the players if
they share the right classical information before they get
separated. However, the game becomes more challenging if the described
correlation is {\em nonlocal}, that is, if it violates a Bell
inequality. In this case any simulation must be
faulty~\cite{Bell-1964}. It is then natural to ask which minimal set of
nonlocal resources the players additionally require to win the game. If
such a set allows for simulating any nonsignaling correlation, it can be
considered a unit of nonlocality. The Popescu-Rohrlich
box~\cite{khalfi85,PR-1994} (PR box) has been shown to be a unit for
bipartite correlations restricted to binary
outputs~\cite{jones-2005,pironio-2005} and, in the converse situation,
for bipartite correlations restricted to binary
inputs~\cite{barret-2005}. In the latter case the simulation can be made
arbitrary close but not perfect. The existence of a bipartite unit that
allows the zero-error simulation of any target correlation, has been
ruled out by counter examples in the multipartite
case~\cite{pironio-2005} and later also for bipartite target
correlations~\cite{dupuis-2007}.

However, the study of a unit of nonlocality is based on the analogous
result in entanglement theory that establishes the singlet as a unit of
entanglement. Some of these transformations are not error-free but
rather approximations that can be made arbitrarily close. It is clear
that in the described simulation game, as well as in real-world
experiments, a sufficiently small simulation error can be hidden from
the tester. Furthermore, in the context of information processing tasks,
asymptotic reductions between different nonlocal resources are often
satisfying. It is therefore both natural and meaningful to establish a
unit that allows for asymptotically perfect simulations.

This is the aim of the present article. The PR box, as suggested by
various contributions, is confirmed to have the properties of an
asymptotic unit of bipartite nonlocality. We will describe a hierarchy
of simulation protocols that allows two players to transform a finite
supply of PR boxes into an arbitrarily close approximation to any
desired bipartite correlation (Theorem \ref{main bipartite
  cor}). Furthermore, we will analyze the simulation's performance in
terms of resource consumption (Theorem \ref{upper bound on the number of
  required PR boxes}). Then, the theoretical possibility of bipartite
units for the multipartite case is studied. We demonstrate two
limitations. Inspired by \cite{pironio-2005}, we construct an instance
of a multipartite simulation game that can provably not be won
arbitrarily often with non-adaptive protocols even when the players have
access to any set of bipartite nonsignaling resources (Corollary \ref{no
  bipartite unit for non-adaptive simulations}). Also, we will show that
if the players are allowed to execute any non-interactive protocol, then
the simulation error of the same game instance is connected with the
balance of the players output functions. As a consequence we calculate a
linear rate at which the simulation distance maximally declines when
increasing the number of shared resource correlations (Theorem \ref{slow
  distance decrease in the general case}).

\section{Preliminaries}\label{pre}

\begin{figure}[h]
  \includegraphics{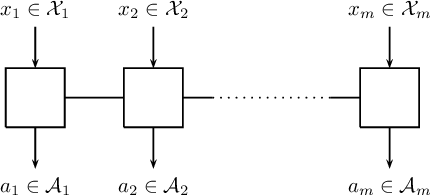}
  \caption{Schematic representation of the ends of an $m$-partite
    system. The system outputs $a=(a_1,a_2,\hdots,a_m)$ on inputs
    $x=(x_1,x_2,\hdots,x_m)$ with probability $P_A^x(a)$.}
  \label{multipartite system}
\end{figure}

Adopting the abstracted approach to nonlocality formulated by {\em
  generalized nonsignaling theories}~\cite{barret-2005}, we consider
correlations in the joint behavior of the ends of an input-output {\em
  system}. An $m$-partite system is characterized by joint probability
distributions $P_A^x$ on $m$ random variables $A=(A_1,\hdots,A_m)$ that
map to the value space $\mathcal{A}_1\times\dots\times \mathcal{A}_m$,
representing the outputs of the system, conditioned on $m$ inputs
$x=(x_1,\hdots,x_m)\in\mathcal{X}_1\times\dots\times\mathcal{X}_m$. See
Figure \ref{multipartite system}.

Furthermore, $P_A^x$ is {\em nonsignaling}, meaning that for any subset
$S\subseteq[m]$ -- we use $[m]$ for short to denote the set
$\{1,\hdots,m\}$ -- the marginal distribution
$P_{A_S}^{x_S,x_{\bar{S}}}$ is independent of the inputs
$x_{\bar{S}}=\{x_i:i\in [m]\backslash S\}$.

Now, consider the following task: In a first phase, a group of $m$
players is given a description of a system $T$, called the {\em target
  system}. Furthermore, the players can share classical information,
such as a global value $\lambda$ from some distribution $P_\Lambda$, and
choose any collection of {\em resource systems} from some predefined
set. After discussing a strategy and sharing the resource systems among
each other, they are separated in order to prevent any
communication. Then the test phase begins. The players are given inputs
$x=(x_1,\hdots,x_m)$, such that each one of them learns only its own
input and has no information on the other inputs. Then each player
determines an output $a_i$ resulting in an overall output string
$a=(a_1,\hdots,a_m)$, such that, after an arbitrary number of
independent rounds, the {\em simulated system} $S$ is as close as
possible to $T$. This means they aim at minimizing the following
measure.

\begin{definition}[simulation distance]\label{distance between systems}
  A simulated system $S$ approximates a target system $T$ with distance
  \begin{align*}
    \delta(S,T)=\max_{x}\frac{1}{2}\sum_{a} |S_A^x(a)-T_A^{x}(a)|.
  \end{align*}
\end{definition}

For fixed inputs $x$ the outputs of the simulated system and the outputs
of the target system are distributed according to $S_A^x$ and $T_A^x$,
respectively. The distance between the two distributions can be
quantified by their total variation distance $1/2\sum_{a}
|S_A^x(a)-T_A^{x}(a)|$. Informally speaking, the simulation distance
$\delta(S,T)$ expresses the worst total variation distance that a tester
may reveal between the simulated system and the target system.

The strategy on which the players agree is called a {\em simulation
  protocol}. It typically includes a plan of which classical
distribution and which resource systems are shared and how they are used
by each one of them to help in the simulation. The players can apply any
classical circuitry to their local parts of the shared systems. Such a
local input-output strategy is called a {\em wiring}~
\cite{barret-2005,wiring2006}. Note that when interacting with a system
one receives an output immediately after providing an input,
independently of whether the player in possession of the other end has
given its input already. This is an allowed convention because all
systems satisfy the nonsignaling constraints.

\subsection{Simulation protocols}

Given inputs $x_1,\hdots,x_m$ and a global random value $\lambda$, which
is drawn from the distribution $P_{\Lambda}$, the players execute their
local protocols on the shared resource systems and determine a final
output. We identify two classes of simulation protocols by
distinguishing the players' local strategies.

The first class allows each player's wiring to consist of arbitrary
local, classical operations on the inputs and outputs of the shared
resource systems. This is the most general description of a simulation
protocol, which we shall call {\em adaptive}. Suppose that player
$i\in[m]$ shares the $n$ resource systems $R_1,\hdots,R_n$ with other
players. In an adaptive simulation protocol, given an input $x_i$ and
the shared random value $\lambda$, player $i$'s possibilities consist of
the following parts.

\begin{enumerate}
\item Player $i$ inputs $f_1(x_i,\lambda)$ to the shared system
  $R_{i_1}$, where the index $i_1$ is determined by the function
  $i_1(x_i,\lambda)$. System $R_{i_1}$ outputs $b_{i_1}$ to player $i$.
\item Player $i$ then inputs $f_2(x_i,\lambda,b_{i_1})$ to $R_{i_2}$,
  where the index of the second system to use is determined by the local
  function $i_2=i_2(x_i,\lambda,b_{i_1})$, obtaining the output
  $b_{i_2}$.
  \begin{center}
    $\vdots$
  \end{center}
\item[n.] Player $i$ continues doing so until all $n$ systems have
  provided an output. In the end the local variables
  $b=(b_1,\hdots,b_n)$ are completely assigned. The final output of
  player $i$ is then given by the result of the local function
  $f^{x_i}(\lambda,b)$.
\end{enumerate}

Locally, an adaptive simulation protocol may consist of as many as $n$
dependent blocks of classical operations, or rounds, generating inputs
to and obtaining outputs from the shared systems.

The second class imposes the natural restriction of parallelism to the
set of adaptive protocols. Each player is limited to execute a {\em
  single} block of classical operations. Thus, this class includes only
those protocols in which each player determines the inputs into all
shared resource systems solely from its initial input and $\lambda$. In
these so-called {\em non-adaptive} simulation protocols the wiring for
player $i$ is such that no input into a resource system depends on the
output of another, and there is no order in using the systems -- all
ends can be evaluated in parallel, immediately after having learned
$x_i$ and $\lambda$. Therefore, any wiring of a non-adaptive protocol
can be described by the input functions
$f_1(x_i,\lambda),\hdots,f_n(x_i,\lambda)$ and the final output function
$f^{x_i}(\lambda,b)$.

In both classes, each player has the freedom to define its own
collection of local functions and, therefore, an individual wiring of
the described kind.

\subsection{Classes of systems}

Typically, an instance of the simulation game is challenging if the set
of resource systems the players are allowed to use, is restricted and
the target correlation is nonlocal. We will define a set of multipartite
target systems and a set of bipartite resource systems for which the
simulation game is particularly difficult.

The following system is inspired by a GHZ quantum correlation (after the
authors of~\cite{greenberger89}, Greenberger, Horne, and Zeilinger)
exhibiting quantum nonlocality~\cite{DiVincenzo-97}. See
also~\cite{pironio-2005}, which introduces this target system in the
present context of simulation games.

\begin{ex}\label{hard target system}
  Let $T$ be any five-partite system with binary inputs
  $\mathcal{X}_1,\hdots,\mathcal{X}_5=\{0,1\}$ and binary outputs
  $\mathcal{A}_1,\hdots,\mathcal{A}_5=\{0,1\}$, fulfilling the following
  six correlation conditions:
  \begin{align}
    \text{If }x_1=0,x_2=1&,x_3=0,\notag\\
    &\text{then }a_1+a_2+a_3\equiv 0~({\text{\em mod }}2).\label{cond1}\\
    \text{If }x_2=0,x_3=1&,x_4=0,\notag\\
    &\text{then }a_2+a_3+a_4\equiv 0~({\text{\em mod }}2).\label{cond2}\\
    \text{If }x_3=0,x_4=1&,x_5=0,\notag\\
    &\text{then }a_3+a_4+a_5\equiv 0~({\text{\em mod }}2).\label{cond3}\\
    \text{If }x_4=0,x_5=1&,x_1=0,\notag\\
    &\text{then }a_4+a_5+a_1\equiv 0~({\text{\em mod }}2).\label{cond4}\\
    \text{If }x_5=0,x_1=1&,x_2=0,\notag\\
    &\text{then }a_5+a_1+a_2\equiv 0~({\text{\em mod }}2).\label{cond5}\\
    \text{If }x_1=x_2=x_3~&=x_4=x_5=1,\notag\\
    \text{then }a_1+&~a_2+a_3+a_4+a_5\equiv 1~({\text{\em mod
      }}2).\label{cond6}
  \end{align}
  For measurements on the corresponding quantum state, we additionally
  have that all output bits, and the parity of all subsets of output
  bits, that are not specified above, are uniformly random.
\end{ex}

Let $\mathcal{P}^b$ stand for the set of all systems with $m=2$,
henceforth called {\em bipartite} systems. The following class of
bipartite resource systems essentially calculates any decision problem
distributed between two parties.

\begin{definition}\label{resource systems}
  Let the set $\mathcal{R}$ include all bipartite systems $R_{AB}^{xy}$
  with binary output alphabets $\mathcal{A}=\mathcal{B}=\{0,1\}$, and
  the joint probability distributions
  \[
  R_ {AB}^{xy}(a,b)=\begin{cases}
    \tfrac{1}{2}&\text{ if }a+b\equiv g(x,y)~({\text{\em mod }}2),\\
    0&\text{ otherwise},
  \end{cases}
  \]
  where $g:\mathcal{X}\times\mathcal{Y}\rightarrow\{0,1\}$ is an
  arbitrary Boolean function on the inputs.
\end{definition}

The next special class of bipartite systems generalizes the class
$\mathcal{R}$ to an arbitrary output alphabet size.

\begin{definition}\label{def: D}
  Let the set $\mathcal{D}$ include all bipartite systems $D_{AB}^{xy}$
  with output alphabets $\mathcal{A}=\mathcal{B}=[d]$, for any integer
  $d>1$, and the joint probability distributions
  \[
  D_{AB}^{xy}(a,b)=
  \begin{cases}
    \tfrac{1}{d}&\text{if }f^{xy}(a)=b,\\
    0&\text{otherwise},
  \end{cases}
  \]
  where $f^{xy}:[d]\re[d]$ are permutations on the output set indexed by
  the inputs $x,y$.
\end{definition}

Every system $D\in\mathcal{D}$ can equivalently be described by input
alphabets $\mathcal{X},\mathcal{Y}$, an integer $d>1$ and a set of
permutations $\{f^{xy}:x\in \mathcal{X},y\in \mathcal{Y}\}$ on $[d]$.

It is easy to see that the marginal probabilities $D_{A}^{x}(a)$ as well
as $D_{B}^y(b)$ are uniform, independently of $y$ and $x$,
respectively. Therefore, $\mathcal{D}\subset\mathcal{P}^b$
holds. Obviously, we have the relationship
$\mathcal{R}\subset\mathcal{D}$ with any system in $\mathcal{R}$
identified by $\mathcal{X},\mathcal{Y}$, the integer $d=2$ and the set
of permutations $\{g(x,y)+a~(\text{mod
}2):x\in\mathcal{X},y\in\mathcal{Y}\}$.

We refer to the PR box~\cite{khalfi85,PR-1994} as the most prominent
element of $\mathcal{R}$. Its correlations can be described as follows:
On inputs $x,y\in\{0,1\}$, the system returns outputs $a,b\in\{0,1\}$,
such that $a$ and $b$ alone are uniform and independent of $(x,y)$, but
the correlation $xy\equiv a+b~(\text{mod }2)$ always holds.

\begin{definition}\label{prbox}
  The PR box is a bipartite system with binary output alphabets
  $\mathcal{X}=\mathcal{Y}=\{0,1\}$ and binary input alphabets
  $\mathcal{A}=\mathcal{B}=\{0,1\}$ and the joint probability
  distributions
  \[
  P_{PR}^{xy}(a,b)=
  \begin{cases}
    \tfrac{1}{2}&\text{if }a+b\equiv xy~({\text{\em mod }}2),\\
    0&\text{otherwise}.
  \end{cases}
  \]
\end{definition}

The PR box violates the Clauser-Horne-Shimony-Holt
inequality~\cite{CHSH-1969} -- a Bell inequality in minimal dimensions
-- to the algebraic maximum. It follows from a work of
Tsirelson~\cite{cirelson-1980} that the correlation of the PR box is
super-quantum, that is, that it cannot be approximated arbitrarily well by
two parties performing quantum mechanical experiments. However, it can
be approximated with an accuracy of roughly $85\%$, whereas $75\%$ is
the local limit.

Finally, we introduce a compact way to describe different simulation
results.

\begin{definition}
  We denote the existence of simulation protocols approximating any
  target system of the set $\mathcal{T}$ with resource systems
  restricted to the set $\mathcal{R}$ by the notation
  $\mathcal{R}\leadsto^\e\mathcal{T}$ and the possibility of zero-error
  simulations by $\mathcal{R}\leadsto\mathcal{T}$.
\end{definition}

We are now sufficiently equipped to start with the statements and proofs
of this articles contributions.

\section{A unit of bipartite nonlocality}\label{bip unit}

In this section, we consider the simulation game for any system in
$\mathcal{P}^b$. Given the description of any bipartite system, we ask
which minimal set of bipartite resource systems is required by two
players to agree on a simulation strategy that imitates the specified
target arbitrarily well. As the main result of this section, we will
prove that a finite supply of copies of a PR box as a resource is
sufficient for this task.

Van Dam~\cite{vandam-2005} has given a construction of arbitrary Boolean
functions distributed between two parties, which coincide with our set
$\mathcal{R}$, with shared PR boxes. We will use this result later. In
an intermediate step, we prove the existence of simulation protocols
approximating $\mathcal{D}$ with resources from $\mathcal{R}$ (Lemma
\ref{first reduction step}). This insight will then be generalized by an
explicit, asymptotic reduction of arbitrary bipartite nonsignaling
systems to $\mathcal{D}$ (Lemma \ref{second reduction step}). The main
result (Theorem \ref{main bipartite cor}) is a consequence of these
three parts as illustrated by the following proof outline.

\renewcommand{\tabcolsep}{0.2cm}
\begin{center}
  \begin{tabular}{ccccc}
    \cite{vandam-2005}&Lemma \ref{first reduction step}&Lemma \ref{second reduction step}&&Theorem \ref{main bipartite cor}
    \\
    &&&&\\
    $P_{PR}\leadsto\mathcal{R}$&$\mathcal{R}\leadsto^\e\mathcal{D}$&$\mathcal{D}\leadsto^\e\mathcal{P}^b$&$\Rightarrow$&$P_{PR}\leadsto^\e\mathcal{P}^b$
    \\
  \end{tabular}  
\end{center}

\subsection{Interconverting $\mathcal{R}$ and $\mathcal{D}$}

As a first step we concentrate on the slightly simpler situation where
the target system is from $\mathcal{D}$, and the resources must be
elements of $\mathcal{R}$. The proof idea is as follows: We argue
inductively, over the size of the output alphabets, by constructing a
simulation protocol that approximates a system $D\in\mathcal{D}$, with
output sets $\mathcal{A}=\mathcal{B}=[d]$, from shared randomness and a
finite number of copies of a certain system $D'(D)\in\mathcal{D}$ with
$\mathcal{A}'=\mathcal{B}'=[d-1]$. Roughly speaking, our protocol uses
appropriately chosen permutations on the smaller output set of the
resource system $D'(D)$ to approximate the permutation distributions of
$D$. For $d=2$ we recover the resource set $\mathcal{R}$ and, therefore,
$\mathcal{R}\leadsto^\e\mathcal{D}$ follows.

Suppose we are given the system $D\in\mathcal{D}$ with input sets
$\mathcal{X}$ and $\mathcal{Y}$ and output alphabet $[d]$. Then, let the
system $D'(D)$ be defined by the set
\[
\{f^{(x,\hat{a})(y,\hat{b})}:(x,\hat{a})\in\mathcal{X}\times[d],(y,\hat{b})\in\mathcal{Y}\times[d]\}
\]
of functions. Each $f^{(x,\hat{a})(y,\hat{b})}:[d-1]\rightarrow[d-1]$ is
constructed from the set of permutations defining $D$, that is from
$\{f^{xy}:x\in\mathcal{X},y\in\mathcal{Y}\}$, by the following rule.
\begin{align}
  f^{(x,\hat{a})(y,\hat{b})}(a)=\left\{
    \begin{array}{ll}
      r_{\hat{b}}^{-1}(f^{xy}(\hat{a}))&\text{if }f^{xy}(a)=\hat{b},a\neq \hat{a},
      \\
      r_{\hat{b}}^{-1}(f^{xy}(r_{\hat{a}}(a)))&\text{otherwise},
    \end{array}
  \right.
  \label{definition of the new permutation}
\end{align}
where $r_{\hat{a}}:[d-1]\re [d]\backslash\{\hat{a}\}$ and
$r_{\hat{b}}:[d-1]\re [d]\backslash\{\hat{b}\}$ are the simple
bijections
\begin{align}
  r_{\hat{a}}(a)=\left\{
    \begin{array}{ll}
      d&\text{if }a=\hat{a},\\
      a&\text{otherwise},
    \end{array}
  \right.~~r_{\hat{b}}(b)=\left\{
    \begin{array}{ll}
      d&\text{if }b=\hat{b},\\
      b&\text{otherwise}.
    \end{array}\right.\label{a0}
\end{align}
For any given $D\in\mathcal{D}$ this construction yields a new system
$D'(D)$ that depends on the permutations defining $D$ and has an output
alphabet that lacks one element compared to $D$. As promised, we can
show that $D'(D)$ is also in $\mathcal{D}$.
\begin{lem}
  If $D\in\mathcal{D}$, then $D'(D)\in\mathcal{D}$.
\end{lem}
\begin{proof}
  We will show that for all $(x,\hat{a})\in\mathcal{X}\times[d]$ and all
  $(y,\hat{b})\in\mathcal{Y}\times[d]$, the function
  $f^{(x,\hat{a})(y,\hat{b})}$ is a permutation on the set $[d-1]$. For
  all inputs $a\in[d-1]$ we have $r_{\hat{a}}(a)\neq \hat{a}$ by
  (\ref{a0}). Therefore, we need only to distinguish two cases from
  (\ref{definition of the new permutation}). First,
  \begin{align*}
    \forall a,a'\in [d-1]:~&f^{(x,\hat{a})(y,\hat{b})}(a)=f^{(x,\hat{a})(y,\hat{b})}(a')\\
    &\Rightarrow r_{\hat{b}}^{-1}(f^{xy}(\hat{a}))=r_{\hat{b}}^{-1}(f^{xy}(\hat{a}))\\
    &\Rightarrow f^{xy}(a)=\hat{b},f^{xy}(a')=\hat{b},\\
    &~~~~~~~~a\neq \hat{a},a'\neq \hat{a}\\
    &\Rightarrow f^{xy}(a)=f^{xy}(a')\\
    &\Rightarrow a=a',
  \end{align*}
  where we used that $D\in\mathcal{D}$ and, therefore, implicitly that
  for any $x,y$ the function $f^{xy}$ is a permutation. Second,
  \begin{align*}
    \forall a,a'\in
    [d-1]:~&f^{(x,\hat{a})(y,\hat{b})}(a)=f^{(x,\hat{a})(y,\hat{b})}(a')
    \\
    &\Rightarrow
    r_{\hat{b}}^{-1}(f^{xy}(r_{\hat{a}}(a)))=r_{\hat{b}}^{-1}(f^{xy}(r_{\hat{a}}(a')))
    \\
    &\Rightarrow a=a',
  \end{align*}
  since $r_{\hat{b}}^{-1}$ and $r_{\hat{a}}$ are defined as bijections
  and we have that $f^{xy}$ is a permutation. Thus,
  $f^{(x,\hat{a})(y,\hat{b})}$ is injective from a finite set to itself
  | a permutation on $[d-1]$.\end{proof}

Now we describe the classical, local operations on which two players,
called Alice and Bob, can agree before they get separated and which will
allow them to emulate any wanted $D\in\mathcal{D}$ from a finite supply
of shared copies of the system $D'(D)\in\mathcal{D}$ to an arbitrary
simulation distance. The simulation protocol consists of a finite number
of rounds. Each round includes {\em four steps} subsequently and locally
executed by Alice and Bob.

\begin{enumerate}
\item The first step is different in the initial round than in
  subsequent rounds.

  In the {\em initial round} Alice draws the local value
  $\hat{a}\sim\mathcal{U}_{[d]}$ and Bob draws
  $\hat{b}\sim\mathcal{U}_{[d]}$, that is, uniformly at random from the set
  $[d]$.

  Otherwise, in any {\em subsequent round} of the simulation, Alice uses
  the already obtained local value $a_t\in[d]$ and Bob uses $b_t\in[d]$,
  respectively, to assign $\hat{a}=a_t$ and $\hat{b}=b_t$.
\item Alice and Bob obtain the shared random bit $\lambda$, such that
  $P_\Lambda(\lambda=0)=1/d$ and $\lambda=1$ otherwise.
\item The shared resource system $D'$ gets inputs $(x,\hat{a})$ from
  Alice and $(y,\hat{b})$ from Bob and outputs $a\in[d-1]$ to Alice and
  $b\in[d-1]$ to Bob.
\item Alice and Bob then process the obtained local data to derive the
  values $a_t\in[d]$ and $b_t\in[d]$ as
  \[
  a_t=
  \begin{cases}
    \hat{a}&\text{if }\lambda=0,\\
    r_{\hat{a}}(a)&\text{otherwise},
  \end{cases}~~b_t=
  \begin{cases}
    \hat{b}&\text{if }\lambda=0,\\
    r_{\hat{b}}(b)&\text{otherwise}.
  \end{cases}
  \]
\end{enumerate}

\begin{figure}[p]
  \includegraphics*[scale=0.98]{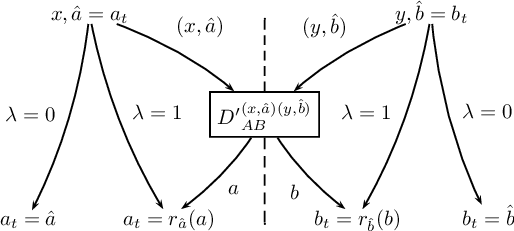}
  \caption{One round of the simulation protocol. On local data $x,a_t$
    and $y,b_t$, Alice and Bob decide to use the outputs of $D'$ on
    inputs $(x,\hat{a})$ and $(y,\hat{b})$ with a probability of
    $P_\Lambda(\lambda=1)$ and to continue with the original pair
    $\hat{a},\hat{b}$ otherwise.}
  \label{one simulation round pic}
  \vspace{1.5cm}
  \includegraphics{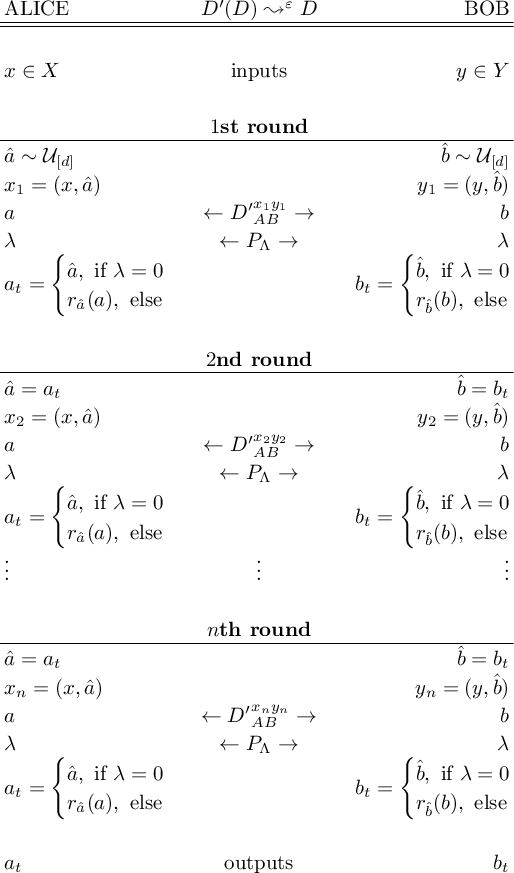}
  \caption{The approximate simulation of $D$ by shared randomness and
    local operations on $n$ copies of the system $D'(D)$.}
  \label{simulation of D pic}
\end{figure}

The next round of the protocol starts with the first step as described
above and proceeds according to the second step and so on. Figure
\ref{one simulation round pic} illustrates a subsequent round. After the
last round the parties output $a_t$ and $b_t$ as the final outputs of
the simulation.

The relabeling functions $r_{\hat{a}}$, $r_{\hat{b}}$ are necessary
because the set $[d-1]$, on which the permutation
$f^{(x,\hat{a})(y,\hat{b})}$ is defined, obviously lacks the element
$d$, which is one of $D$'s outputs. On the other hand, we have already
correlated the outputs $\hat{a}$ and $\hat{b}$ in the case $\lambda=0$,
so this pair of outputs can serve as a substitution for $d$ and
$f^{xy}(d)$, respectively.

Figure \ref{simulation of D pic} illustrates the protocol approximating
a target system $D$ with $n$ rounds.

Now we prove two lemmas that describe useful properties of the presented
protocol. Given are inputs $x,y$. At the beginning of every round, Alice
and Bob hold a pair $\hat{a},\hat{b}$. If $f^{xy}(\hat{a})\neq\hat{b}$,
then we have a certain probability with which the simulation of $D$
fails in that round.

\begin{lem}\label{corr1}
  In a round initialized with $\hat{a}$ and $\hat{b}$, such that
  $f^{xy}(\hat{a})\neq\hat{b}$, Alice and Bob generate outputs $a_t$ and
  $b_t$ that do not agree with the required correlation in $D_{AB}^{xy}$
  with probability $2/d$.
\end{lem}

\begin{proof} After initializing $\hat{a}$ and $\hat{b}$ the second step
  of the simulation round follows. The possible events are:
  \medskip\\
  If $\lambda=0$, where $P_\Lambda(0)=1/d$, then Alice and Bob
  assign $a_t=\hat{a},b_t=\hat{b}$, which is an incorrect correlation.
  \medskip\\
  If $\lambda=1$, where $P_\Lambda(1)=(d-1)/d$, then Alice and Bob
  assign $a_t=r_{\hat{a}}(a)$, $b_t=r_{\hat{b}}(b)$. By (\ref{definition
    of the new permutation}), the definition of
  $f^{(x,\hat{a})(y,\hat{b})}$, we have that the output pair $a,b$
  obtained from $D'$ obeys the permutation
  $f^{(x,\hat{a})(y,\hat{b})}(a)=b$. In this case we must distinguish
  three possible situations:
  \medskip\\
  (1) If Alice gets $a$ such that $f^{xy}(a)=\hat{b}$, which happens
  with a maximal probability of $1/(d-1)$, then an error occurs
  because, by the definition of $r_{\hat{a}}$, Alice will never output
  $\hat{a}$ if $\lambda=1$. The round then finishes with the pair
  \begin{align*}
    a_t&=r_{\hat{a}}(a)=a,\\
    b_t&=r_{\hat{b}}(f^{(x,\hat{a})(y,\hat{b})}(a))=r_{\hat{b}}(r_{\hat{b}}^{-1}(f^{xy}(\hat{a})))=f^{xy}(\hat{a})\neq
    \hat{b}.
  \end{align*}
  (2) If Alice gets $a=\hat{a}$, which can happen if $\hat{a}\neq d$,
  then the round finishes with the correctly correlated pair
  \begin{align*}
    a_t&=r_{\hat{a}}(\hat{a})=d,\\
    b_t&=r_{\hat{b}}(f^{(x,\hat{a})(y,\hat{b})}(\hat{a}))=r_{\hat{b}}(r_{\hat{b}}^{-1}(f^{xy}(r_{\hat{a}}(\hat{a}))))=f^{xy}(d).
  \end{align*}
  (3) If Alice gets any other $a\in[d-1]$, then the round finishes with
  the correctly correlated pair
  \begin{align*}
    a_t&=r_{\hat{a}}(a)=a,\\
    b_t&=r_{\hat{b}}(f^{(x,\hat{a})(y,\hat{b})}(a))=r_{\hat{b}}(r_{\hat{b}}^{-1}(f^{xy}(r_{\hat{a}}(a))))=f^{xy}(a).
  \end{align*}

  \begin{figure}[t]
    \includegraphics{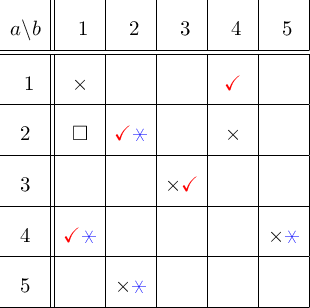}
    \caption{An illustration of Lemma \ref{corr1}: Suppose a round
      initialized with $\hat{a}=2$ and $\hat{b}=1$ such that
      $f^{xy}(\hat{a})\neq\hat{b}$ ($\square$). Such a permutation on
      the set $[d=5]$ is indicated with crosses ($\times$). The
      corresponding permutation $f^{(x,\hat{a})(y,\hat{b})}$ is shown
      with check-marks (\textcolor{red}{\checkmark}). The relabelings
      $r_{\hat{a}=2}$ and $r_{\hat{b}=1}$ correlate $(5,2)$ by $(2,2)$
      and $(4,5)$ by $(4,1)$ (indicated by
      \textcolor{blue}{$\hexstar$}). Therefore, this round ends with an
      incorrect pair if $\lambda=0$ or if Alice gets $a=1$, which
      happens with a total chance $2/5$.}
    \label{bad round example}
  \end{figure}

  Therefore, the round will certainly end in a bad pair if
  $\lambda=0$. Otherwise, if $\lambda=1$, at most one pair of outputs
  from $D'$ yields an incorrect correlation. We get an overall
  probability for a final pair $a_t,b_t$, that does not satisfy the
  permutation $f^{xy}$, of
  $P_\Lambda(0)+P_\Lambda(1)\cdot 1/(d-1)=2/d$.\end{proof}

See Figure \ref{bad round example} for an illustration of Lemma
\ref{corr1}. The next lemma assures, that once a pair $a_t,b_t$
satisfying $f^{xy}(a_t)=b_t$, is found, all following rounds will
simulate exactly the distribution $D^{xy}_{AB}$.

\begin{lem}\label{corr2}
  In a round initialized with $\hat{a}$ and $\hat{b}$, such that
  $f^{xy}(\hat{a})=\hat{b}$, Alice and Bob generate outputs $a_t$ and
  $b_t$ that agree with the required distribution $D_{AB}^{xy}$.
\end{lem}

\begin{proof} After initializing $\hat{a}$ and $\hat{b}$ the second step
  of the simulation round follows. The possible events are:
  \medskip\\
  If $\lambda=0$, where $P_\Lambda(0)=1/d$, then Alice and Bob
  assign $a_t=\hat{a},b_t=\hat{b}$, which is a correct pair.
  \medskip\\
  If $\lambda=1$, where $P_\Lambda(1)=(d-1)/d$, then Alice and Bob
  assign $a_t=r_{\hat{a}}(a)$, $b_t=r_{\hat{b}}(b)$. By (\ref{definition
    of the new permutation}) we have that the output pair $a,b$ obtained
  from $D'$ obeys the permutation $f^{(x,\hat{a})(y,\hat{b})}(a)=b$. We
  must distinguish two situations:
  \medskip\\
  (1) If Alice gets $a=\hat{a}$, which could happen with probability
  $1/(d-1)$ if $\hat{a}\neq d$, then the round finishes with the
  correctly correlated pair
  \begin{align*}
    a_t&=r_{\hat{a}}(\hat{a})=d,\\
    b_t&=r_{\hat{b}}(f^{(x,\hat{a})(y,\hat{b})}(\hat{a}))=r_{\hat{b}}(r_{\hat{b}}^{-1}(f^{xy}(r_{\hat{a}}(\hat{a}))))=f^{xy}(d).
  \end{align*}
  (2) If Alice gets any other $a$, then the round finishes with the
  correctly correlated pair
  \begin{align*}
    a_t&=r_{\hat{a}}(a)=a,\\
    b_t&=r_{\hat{b}}(f^{(x,\hat{a})(y,\hat{b})}(a))=r_{\hat{b}}(r_{\hat{b}}^{-1}(f^{xy}(r_{\hat{a}}(a))))=f^{xy}(a).
  \end{align*}
  
  \begin{figure}[t]
    \includegraphics{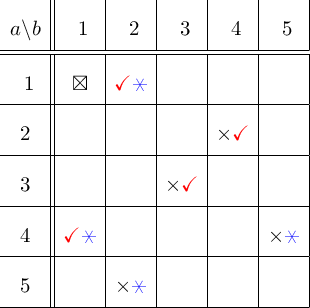}
    \caption{An illustration of Lemma \ref{corr2}: Suppose a round
      initialized with $\hat{a}=1$ and $\hat{b}=1$ such that
      $f^{xy}(\hat{a})=\hat{b}$ ($\square$). Such a permutation on the
      set $[d=5]$ is indicated with crosses ($\times$). The
      corresponding permutation $f^{(x,\hat{a})(y,\hat{b})}$ is shown
      with check-marks (\textcolor{red}{\checkmark}). The relabelings
      $r_{\hat{a}=1}$ and $r_{\hat{b}=1}$ correlate $(5,2)$ by $(1,2)$
      and $(4,5)$ by $(4,1)$ (indicated by
      \textcolor{blue}{$\hexstar$}). Therefore, this round ends with
      only correct pairs that occur uniformly at random.}
    \label{good round example}
  \end{figure}
  
  Thus, the round establishes local values $a_t,b_t$ for which
  $f^{xy}(a_t)=b_t$ holds. It is easy to see that each pair is equally
  probable, that is, happens with probability $1/d$. Therefore the
  round reproduces the joint distribution $D^{xy}_{AB}$
  correctly.\end{proof}

See Figure \ref{good round example} for an illustration of Lemma
\ref{corr2}. We now prove that the presented protocol achieves an
arbitrarily good simulation of any $D\in\mathcal{D}$ using a finite
supply of approximations to $D'(D)$, which are not too faulty.

\newpage

\begin{lem}\label{first reduction step}
  It holds that $\mathcal{R}\leadsto^\e \mathcal{D}$.
\end{lem}

\begin{proof} According to the simulation game setup two players are
  given the description of any target system in $\mathcal{D}$ with
  output alphabet $[d]$, where $d>2$. We denote this target system with
  $D_d$. The two players have a supply of $n$ approximations to the
  system $D'(D_d)$ at their disposal. By executing the presented
  protocol for $D_d$ they simulate the system $S$ as shown in Figure
  \ref{simulation of D pic}. We measure the quality of their
  approximation by the distance $\delta(S,D_d)$, as introduced in
  Definition \ref{distance between systems}.

  If the players use $n$ error-free resource systems of the kind $D'$,
  then, in each round, Lemmas \ref{corr1} and \ref{corr2} imply a
  probability of $(d-2)/d$ to reach a zero-error simulation of
  $D_d$. Therefore, the probability that a wrong correlation remains
  after $n$ subsequent rounds is at most
  $(2/d)^n$. So, we have
  \begin{align}
    \delta(S,D_d)<\left(\frac{2}{d}\right)^{n},
    \label{delta upper bound 1}
  \end{align}
  where we obtain a strict upper bound because we omitted the non-zero
  probability to guess a correct pair in the first round. However, if
  the players feed the simulation protocol with $n$ approximations
  (denoted $S_{d-1}$) to the resource $D_{d-1}=D'$, the simulation error
  $\delta_{d-1}=\delta(S_{d-1},D_{d-1})$ needs to be taken into
  consideration.

  In each round the current simulated resource system $S_{d-1}$ returns
  a pair of outcomes not according to its definition with a chance of at
  most $\delta_{d-1}$. In this case the round can terminate with a wrong
  pair. If the resource system returns a pair to the players as its
  specification $D_{d-1}$ dictates, then a wrong pair remains with
  probability $2/d$. Thus, on inputs $x,y$, an incorrectly
  initialized round does not succeed in simulating the distribution
  $D_{AB}^{xy}$ with probability at most
  $\delta_{d-1}+(1-\delta_{d-1})2/d$ (see Figure \ref{one
    simulation round with faulty resources}, left-hand side). A round
  starting with a correct pair can still be influenced by the faulty
  simulation of $D_{d-1}$. With a maximal probability of $\delta_{d-1}$
  the initially correct pair gets corrupted (see Figure \ref{one
    simulation round with faulty resources}, right-hand side).

  \begin{figure}[h]
    \includegraphics{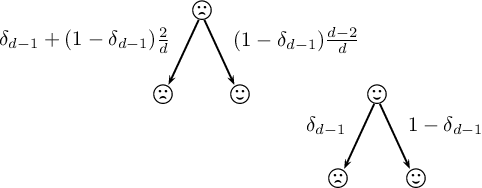}
    \caption{In the left tree are the transition probabilities if the
      round starts with a faulty pair ({\frownie}) and in the tree on
      the right side we illustrate the transitions in the case of a
      correct starting pair.}
    \label{one simulation round with faulty resources}
  \end{figure}

  Let $\delta^{(n)}_d$ denote the final simulation error if we run the
  simulation protocol for $D_d$ on exactly $n>0$ resource systems, that is,
  for $n$ subsequent rounds. Then, for any $d>2$, one can derive the
  recursive formula
  \begin{align}
    \delta^{(n)}_d&<\delta_d^{(n-1)}\left(\delta_{d-1}+(1-\delta_{d-1})\frac{2}{d}\right)+(1-\delta^{(n-1)}_d)\delta_{d-1}\notag
    \\
    &=\delta^{(n-1)}_d(1-\delta_{d-1})\frac{2}{d}+\delta_{d-1}
    \label{recursive distance formula}
  \end{align}
  with the trivial base cases $\delta_2=0$ and $\delta^{(0)}_d=1$. For
  the simplicity of the formula we assumed that the initial pair is
  wrong -- ignoring the fact that with some non-zero probability the
  initial guess is correct -- which results in a strict upper bound. The
  explicit form of the right hand side of (\ref{recursive distance
    formula}) can be found easily and turns out to be
  \begin{align}
    \delta^{(n)}_d<\left(\frac{2}{d}(1-\delta_{d-1})\right)^n+\delta_{d-1}\left(\frac{\left(\frac{2}{d}(1-\delta_{d-1})\right)^n-1}{\frac{2}{d}(1-\delta_{d-1})-1}\right)
    \label{delta with n resources}
  \end{align}
  with the straightforward limiting expression
  \begin{align}
    \lim_{n\re\infty}\delta^{(n)}_d<\frac{d\delta_{d-1}}{d-2+2\delta_{d-1}},
    \label{limit on the minimal distance}
  \end{align}
  for large $n$. This expresses the intuitive fact that the distance of
  the actual simulated resource systems from their specification
  restricts the success of the shown simulation protocol.

  We will now show that there always exists a finite number of
  resources, departing from their specifications with a certain non-zero
  distance $\delta_{d-1}$, that suffice for simulating $D_d$ to any
  desired quality $\delta_d>\lim_{n\re\infty}\delta^{(n)}_d$.

  First, it follows from (\ref{delta with n resources}) that the
  simulation of any given system $D_3\in\mathcal{D}$ within the maximal
  distance $\delta_3=(2/3)^{n_3}$ can be achieved
  with $n_3$ copies of the resource system
  $D'(D_3)\in\mathcal{R}$. Generally, it is convenient for the following
  analysis and in accordance with the limit (\ref{limit on the minimal
    distance}), to choose, for any $d>3$, the number of rounds as
  \begin{align}
    n_d=\lceil\log_{\frac{2}{d}(1-\delta_{d-1})}\delta_3\rceil,\label{required
      number of rounds}
  \end{align}
  where obviously $\delta_2=0$. With this choice of $n_d$ the upper
  bound on the distance $\delta_d=\delta_d^{(n_d)}$ becomes
  \begin{align}
    \delta_d<\delta_3+\frac{\delta_{d-1}d(1-\delta_3)}{d-2+2\delta_{d-1}}.
    \label{recursive formula for the distance}
  \end{align}
  This recursive distance bound yields, after some simplifications, the
  explicit bound
  \begin{align}
    \delta_d<\delta_3d^2,
    \label{explicit distance after d-2 steps}
  \end{align}
  which is greater or equal to $\lim_{n\re\infty}\delta^{(n)}_d$ if
  $\delta_{d-1}\leq 1$. So, an initial distance of
  \begin{align}
    \delta_3=\frac{\delta_d}{d^2}\label{maximal intial distance formula}
  \end{align}
  is sufficient to guarantee a final distance below
  $\delta_d$. Therefore, for an arbitrary $\delta_d>0$ and an alphabet
  size $d>2$, we calculate $\delta_3$ by (\ref{maximal intial distance
    formula}). From (\ref{recursive formula for the distance}) and
  (\ref{explicit distance after d-2 steps}), with fixed values
  $\delta_d$ and $\delta_3$, the existence of a sufficiently small,
  non-zero distance $\delta_{d-1}$ is implied.

  We conclude that, for any desired $\delta_d$ with $d>2$, there is
  always a finite number $n_d$ -- given by (\ref{required number of
    rounds}) -- of approximations to the resource system $D'(D_d)$ with
  a certain non-zero distance $\delta_{d-1}$, such that our protocol
  achieves the simulation of $D_d$ in the required quality. Since any
  finite amount of all $D_2\in\mathcal{R}$ is available in perfect
  quality to the players, the statement in the lemma follows by
  induction.\end{proof}

We finish with a performance estimate of the shown simulation in terms
of the targeted distance $\delta_d$, for any $d>2$, and the number of
needed resource systems from $\mathcal{R}$. To reach the system $D_d$
the simulation protocol is used $d-2$ times to increase the size of the
output alphabet from 2 to $d$ in each step by 1. In this stepwise
procedure the simulation error grows according to (\ref{recursive
  formula for the distance}) and reaches a distance bounded as in
(\ref{explicit distance after d-2 steps}) after $d-2$ steps. With
\begin{align}
  \prod_{i=3}^dn_i\leq
  \left\lceil\log_{\frac{2}{3}}\delta_3\right\rceil^d=\left\lceil\log_{\frac{3}{2}}\frac{d^2}{\delta_d}\right\rceil^d
  \label{upper bound on the number of resources}
\end{align}
resources from $\mathcal{R}$ our protocol thus simulates a system with a
maximal distance of $\delta_d>0$ from the target $D_d$.

\subsection{Interconverting $\mathcal{D}$ and $\mathcal{P}^b$}

It is the purpose of the following part to close the gap between the set
of permutation systems $\mathcal{D}$ and the set of all bipartite
systems $\mathcal{P}^b$.

We show that two players, having the complete set $\mathcal{D}$
available as a resource, can simulate any bipartite system to an
arbitrarily small distance. For any given $P\in\mathcal{P}^b$ we will
construct a single resource $D(P)\in\mathcal{D}$ on which Alice and Bob
can perform this task.

\begin{lem}\label{second reduction step}
  It holds that $\mathcal{D}\leadsto^\e\mathcal{P}^b$.
\end{lem}

\begin{proof} 
  Suppose we are given a target system $P\in\mathcal{P}^b$ with input
  alphabets $\mathcal{X}$, $\mathcal{Y}$ and output alphabets
  $\mathcal{A}$, $\mathcal{B}$. The task is to find an integer $d$ and a
  system $D(P)\in\mathcal{D}$ with the output set $[d]$, or
  equivalently, a set of permutations $\{f_{xy}:x\in X,y\in Y\}$ on
  $[d]$, such that, for all inputs $x,y$, the distribution $P^{xy}_{AB}$
  can be reproduced by local operations on $D(P)$.
  
  If $P$ contains any irrational probabilities, then we proceed with an
  arbitrarily close approximation to $P$ that is defined by rational
  probabilities only. In this special case the simulation will
  necessarily deviate from the specification and therefore the general
  result $\mathcal{R}\leadsto^\e P$ holds. Otherwise, we can construct
  exact simulations, that is, we can prove the reduction
  $\mathcal{R}\leadsto P$ for entirely rational target systems.

  We construct $D(P)\in\mathcal{D}$ as follows. First, we choose $d$ to
  be the {\em least common denominator} (lcd) of all probabilities
  described by the system $P$. It is calculated as the smallest positive
  integer that is a multiple of all denominators. Therefore, we define
  \[
  d:=\text{lcd}
  \{P_{AB}^{xy}(a,b):x\in\mathcal{X},y\in\mathcal{Y},a\in\mathcal{A},b\in\mathcal{B}\}.
  \]
  Second, we fix local relabelings for given inputs $x,y$, that map all
  pairs $(a,b)$, identified through $f^{xy}(a)=b$, to a certain
  correlation according to $P^{xy}_{AB}$. For any given $x$ let Alice's
  local relabeling function be denoted by
  \[
  r^x_A:[d]\re\mathcal{A}.
  \]
  On input $x$, $r^x_A$ maps, for each $a\in\mathcal{A}$, exactly
  $dP_A^x(a)$ unique elements of the set $[d]$ (the output set of
  $D(P)$) to the value $a$. And similarly for Bob. For any given input
  $y$ let the function
  \[
  r^y_B:[d]\re\mathcal{B},
  \]
  denote Bob's local relabeling of outputs from $D(P)$. On input $y$,
  $r^y_B$ maps, for each $b\in\mathcal{B}$, exactly $dP_B^y(b)$ unique
  elements of the set $[d]$ to the value $b$. Once the relabelings
  $r_A^x,r_B^y$ are fixed for all $x,y$, we can directly derive matching
  permutations $f^{xy}$, for any $x,y$, such that the corresponding
  system $D(P)$ is transformed into $P$ by the protocol consisting of
  the predefined local relabelings of outputs (this idea is illustrated
  in Figure \ref{local relabeling example}).

  \begin{figure}[ht]
    \includegraphics*[scale=0.86]{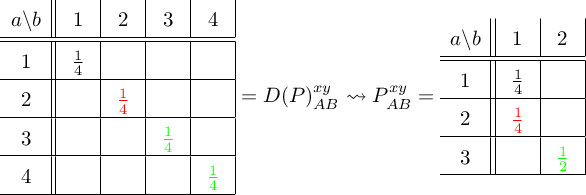}
    \caption{Here, we illustrate the simulation of an example
      distribution $P_{AB}^{xy}$ by local relabelings on the outputs of
      the distribution $D(P)^{xy}_{AB}$ when given fixed inputs
      $x,y$. Alice relabels local outputs from $D(P)^{xy}_{AB}$ as
      $r_A^x(1)=1,r_A^x(2)=2,r_A^x(3)=3$ and $r_A^x(4)=3$. Similarly,
      Bob relabels according to $r_B^y(1)=1,r_B^y(2)=1,r_B^y(3)=2$ and
      $r_B^y(4)=2$. Doing so, the players reproduce $P_{AB}^{xy}$
      exactly.}
    \label{local relabeling example}
  \end{figure}

  Suppose fixed inputs $x,y$. For each pair
  $(a,b)\in\mathcal{A}\times\mathcal{B}$ in the support of
  $P^{xy}_{AB}$, let $f^{xy}$ correlate a number of $dP^{xy}_{AB}(a,b)$
  unique outcomes $a',b'\in[d]$, that is, $f^{xy}(a')=b'$, such that
  $r_A^x(a')=a$ and $r_B^y(b')=b$. Since each output is correlated only
  once, it follows directly that $f^{xy}$ is a permutation and that,
  after the relabeling, all probabilities $P^{xy}_{AB}(a,b)$ are
  recovered. Therefore, we can simulate any wanted $P$ to an arbitrary
  precision by a simulation protocol on the system
  $D(P)\in\mathcal{D}$.\end{proof}

\subsection{Interconverting $P_{PR}$ and $\mathcal{P}^b$}

Putting the three pieces together we are now ready to state the main
theorem of this section.

\begin{thm}\label{main bipartite cor}
  It holds that $P_{PR}\leadsto^\e\mathcal{P}^b$.
\end{thm}

\begin{proof}
  From a finite supply of PR boxes any system in $\mathcal{R}$ can be
  simulated perfectly~\cite{vandam-2005}. We described protocols on
  transforming a finite quantity of copies of a corresponding resource
  system in $\mathcal{R}$ into any system in $\mathcal{D}$ up to any
  desired maximal simulation distance (Lemma \ref{first reduction
    step}). Finally, any bipartite system can be simulated by Alice and
  Bob using a certain system from $\mathcal{D}$ and local operations
  (Lemma \ref{second reduction step}). The final simulation is not
  error-free if the target system contains irrational
  probabilities. Concluding, given enough PR boxes to the disposal of
  Alice and Bob, the two can agree on a classical strategy such that any
  wanted bipartite system is approximated within an arbitrarily small
  distance from its specification.
\end{proof}

We finish the section with some remarks concerning the efficiency of the
shown procedure. Van Dam's construction implies that for a target system
in $\mathcal{R}$ with input alphabets $\mathcal{X}$ and $\mathcal{Y}$,
where w.l.o.g. $|\mathcal{X}|\geq|\mathcal{Y}|$, one requires a supply
of at most $|\mathcal{X}|$ PR boxes. It follows from our reduction
$\mathcal{R}\leadsto^\e\mathcal{D}$ that for simulating a target system
$D_d\in\mathcal{D}$, with output alphabet size $d$ and input sets
$\mathcal{X}$ and $\mathcal{Y}$, we require a number of systems from
$\mathcal{R}$ with input set cardinalities $d!|\mathcal{X}|/2$
and $d!|\mathcal{Y}|/2$. Therefore, using (\ref{upper bound on
  the number of resources}), the simulation of $D_d$, with a maximal
input cardinality of $|\mathcal{X}|$, to the maximal distance
$\delta_d$, costs at most
\[
\left\lceil\log_{\frac{3}{2}}\frac{d^2}{\delta_d}\right\rceil^d\cdot\frac{d!}{2}|\mathcal{X}|
\]
PR boxes. The final reduction $\mathcal{D}\leadsto^\e\mathcal{P}^b$ is a
one to one relation. We simulate any $P\in\mathcal{P}^b$ with a related
system $D_d(P)\in\mathcal{D}$ to a distance
$\delta\leq\delta_d+(1-\delta_d)\delta_I$. Here, $\delta_I$ stands for
the error probability implied by the replacement of irrational
probabilities in $P$ with rational approximations. Therefore, with the
input cardinality $|\mathcal{X}|$ untouched by this reduction, we get
that:
\begin{thm}\label{upper bound on the number of required PR boxes}
  Any system $P\in\mathcal{N}_2$, with a maximal input set
  $\mathcal{X}$, can be simulated by two separated players to any distance $\delta>0$ with
  \[
  \left\lceil\log_{\frac{3}{2}}\frac{d^2(1-\delta_I)}{(\delta-\delta_I)}\right\rceil^d\cdot\frac{d!}{2}|\mathcal{X}|
  \]
  copies of a PR box as shared resources.
\end{thm}
The choice for $\delta_I<\delta$ is arbitrary but implies a
minimal size for the parameter $d$ depending on the actual irrational
probabilities that need to be approximated. Roth's
inequality~\cite{roth55:_ration}, an important result in the field of
Diophantine approximations, provides such a lower
bound~\footnote{{R}oth's famous result on the rational approximation of
  real numbers states that for any irrational algebraic number $a$ of
  degree at least $2$ and any $\e>0$, there is a constant $c(a,\e)$ such
  that $|a-p/q|>c(a,\e)/q^{2+\e}$ holds for any rational
  number $p/q$.}.

None of the shown simulations claims to be optimal in the consumption of
resources, neither does Van Dam's construction. To our present knowledge
it is an open question if the simulation can be improved to run on a
number of PR boxes which is at most exponential in~$d$.

\section{A bipartite unit of nonlocality}\label{multipartite section}

As a natural follow-up we now extend the search for a unit to the space
of target systems with more than two ends. Here, we investigate if, as
it is the case in the bipartite simulation setting, there exists a set
of bipartite resource systems such that a number of separated players
can simulate any desired multipartite nonsignaling correlation
arbitrarily well.

First, we introduce some additional definitions that will be used
throughout this section.

\begin{definition}
  Let a (partial) assignment of values to a set of $n$ variables be
  denoted by $\rho\in(\mathbb{R}\cup \{*\})^n$, where $\rho[v]=*$ means
  that the variable $v$ remains unassigned by the partial assignment
  $\rho$.
\end{definition}

If convenient we will sometimes write $\rho(V)$ for any assignment of
values to the specified set $V$ of variables, while leaving the rest
unassigned ($*$). An example: For $n=3$ we understand the assignment
$\rho=(0,1,*)$ as the mapping from the variables $(v_1,v_2,v_3)$ to
$(0,1,*)$, that is, $v_1=0, v_2=1$ and $v_3$ unassigned. Equivalently,
one may write $\rho(v_1,v_2)=(0,1)$.

\begin{definition}
  For any (partial) assignment $\rho$ let $l(\rho)$ denote the number of
  variables that are fixed by $\rho$. Formally,
  \[
  l(\rho):=|\{v:\rho[v]\neq *\}|.
  \]
\end{definition}

Suppose the inputs to some function $f:X\rightarrow Y$ are drawn from a
probability distribution.

\begin{definition}\label{p}
  For any (partial) assignment $\rho$ of values to the inputs of $f$ let
  $f|_{\rho}$ denote the random variable for the result of $f$ under the
  condition $\rho$. The probability distribution for all $y\in Y$ is
  straightforwardly given by
  \[
  P(f|_\rho=y)=P(f=y|\rho).
  \]
\end{definition}

\begin{definition}
  For any Boolean function $f$ and any (partial) input assignment
  $\rho$, let $\mu(f|_\rho)$ denote the probability that $f$ evaluates
  to the minority decision conditioned on the assignment
  $\rho$. Formally, we define
  \begin{align*}
    \mu(f|_\rho):=\min(P(f|_{\rho}=0),P(f|_{\rho}=1)).
  \end{align*}
\end{definition}

For the next definition we suppose a fixed adaptive simulation protocol
in which an arbitrary player $j\in[m]$ shares $n$ resource systems
$R_1,...,R_n$ with the rest of the players.

\begin{definition}\label{definition of the assignment set}
  For any input $x_j$ and any index $i\in[n]$, let the set $S_i(x_j)$
  include all those partial assignments $\rho$ to the shared variable
  $\lambda$ and the binary variables $b_1,\hdots,b_n$, that, according
  to player $j$'s wiring, imply $i$ as the index of the next system to
  use. Formally, for any $i$, we define
  \[
  S_i(x_j):=\{\rho\in\mathbb{R}\times\{0,1,*\}^n:i_{l(\rho)}(x_j,\rho)=i\}.
  \]
\end{definition}

We can now start with the statements and proofs of the present section.

\subsection{Non-adaptive protocols}

In what follows, a counter example is derived, that is, a simulation
game that cannot be played arbitrarily well under certain
constraints. We consider the simulation of the target system $T$
(Example \ref{hard target system}). The five involved players are
restricted to agree on non-adaptive simulation protocols only and have
the resource set $\mathcal{P}^b$ available. As the main result of this
subsection, it is shown in Corollary \ref{no bipartite unit for
  non-adaptive simulations} that the minimal departure $\delta(S,T)$ of
any system $S$, simulated under these conditions, from the target $T$,
is bounded away from 0.

We use the following notation: Player $i$ shares $n$ resource systems in
total and of these $n_{j}\leq n$ resource systems $R_1,\hdots,R_{n_j}$
with player $j$ in particular. During the simulation game, player $i$ is
given input $x_i$ and the shared random value $\lambda$, drawn from some
distribution $P_\Lambda$. The outputs of the resource systems shared
with player $j$ are assigned to the local variables
$b_1,\hdots,b_{n_j}$, which are part of the overall local sequence
$b_1,\hdots,b_n$. The output of player $i$ is determined by
$f^{x_i}(\lambda,b_1,\hdots,b_n)$.

For Lemma \ref{minimal simulation distance on input 0} we allow only
simulation protocols of a special kind and generalize the results later
in Corollary \ref{no bipartite unit for non-adaptive simulations}. We
suppose any simulation protocol for $T$ executed by five players on
resources from $\mathcal{R}$ with the following {\em input-dependence
  constraint}: Each player determines the inputs into systems shared
with player $j$ independently of the outputs obtained from systems
shared with player $j'$, for all distinct $j,j'\in[5]$.

Lemma \ref{minimal simulation distance on input 0} expresses the fact
that the dependence of a player's output on the local outputs of the
resource systems shared with another player implies a violation of $T$'s
correlation conditions (stated in Example \ref{hard target system}) and,
therefore, a certain related simulation distance. It is an adaptation of
Theorem 2 by Barrett and Pironio~\cite{pironio-2005} to approximate
reductions.

\begin{lem}\label{minimal simulation distance on input 0}
  Take any pair of distinct players $i,j\in[5]$. Any simulated system
  $S$ departs from the target $T$ with distance
  \[
  \delta(S,T)\geq\sum_{\rho}P(\rho)\mu(f^0|_\rho),
  \]
  where $f^0$ is player $i$'s output function on input $0$ and we sum
  over all partial assignments $\rho=\rho(\lambda,b_{n_j+1},\hdots,b_n)$
  to player $i$'s local variables.\end{lem}

\begin{proof}
  We prove the statement for an exemplary pair of players. The reasoning
  extends to any other pair by a simple argument, as explained later.

  Let us choose the pair $i=1,j=2$ and analyze the situation from player
  1's perspective on input $x_1=0$. Note that from the definition of the
  target system and the resources it follows that the sequence
  $b_1,\hdots,b_n$ is a uniformly distributed random binary string of
  length $n$ and $f^0$ -- player 1's local output function on input 0 --
  is a Boolean function.

  Since $x_1=0$, we have that if $x_4=0$ and $x_5=1$, then correlation
  (\ref{cond4}) of Example \ref{hard target system}, that is,
  \begin{align}
    a_4+a_5+a_1\equiv 0~(\text{mod }2)\label{first correlation example}
  \end{align}
  needs to be fulfilled by the final outputs of the players. Conditioned
  on any assignment $\rho=\rho(\lambda,b_{n_2+1},\hdots,b_n)$, the
  result of $f^0$ depends solely on the remaining variables
  $b_1,\hdots,b_{n_2}$. In other words, it depends on the outcomes of
  systems shared with player 2, who is not involved in the above
  correlation. If player 2's actions on these systems would influence
  the simulation of (\ref{first correlation example}), then players 1,4,
  and 5 could team up and receive signals from player 2. Since
  $\mathcal{R}\subset\mathcal{P}$ holds, this is impossible. We can thus
  assume that player 2 does not provide inputs to the systems shared
  with player 1 while sustaining the simulation of (\ref{first
    correlation example}). Therefore, conditioned on $x_1=0$ and any
  $\rho$, player 1's output is basically a local random bit that is 1
  with probability $P(f^0|_\rho=1)$.

  Let $\rho_i$ denote an assignment of values to all outputs of resource
  systems that are received by players $[5]\backslash\{i\}$. If we fix
  their inputs, the shared random value $\lambda$ and $\rho_i$, then the
  outputs of these players are determined.

  Suppose now fixed inputs $x_1=0,x_2,x_3,x_4=0,x_5=1$ and a fixed
  shared value $\lambda$. Conditioned on any $\rho_1$, player 1's
  correct output is uniquely given by (\ref{first correlation example}),
  the other output implies a violation of this correlation. For any
  $\rho$ and $\rho_1$, (\ref{first correlation example}) is thus
  violated with a probability of at least $\mu(f^0|_\rho)$. In all
  protocols considered here player 1's inputs into systems
  $R_{n_2+1},\hdots,R_n$ are independent of
  $b_1,\hdots,b_{n_2}$. Therefore, the distribution of the random
  variable $f^0|_\rho$ is independent of $\rho_1$. The convex
  combination of all possible events $\rho$ and $\rho_1$ yields
  \begin{align*}
    \delta(S,T)&\geq
    \sum_{\rho,\rho_1}P(\rho,\rho_1)\mu(f^0|_\rho)\\
    &=\sum_{\rho}P(\rho)\mu(f^0|_\rho).
  \end{align*}
  This argument extends to any pair of players because one can always
  find a correlation among (\ref{cond1})-(\ref{cond6}) in which only one
  of the players is involved on input 0.
\end{proof}

It is now clear that using resource systems from $\mathcal{R}$ in a
simulation of $T$ guarantees a distance related to the players' output
functions. When a player only considers its initial input and the shared
random value for a final output decision, we expect the simulation distance to be higher. However, this increase might still be
within a constant factor of the lower bound derived above. This is the
idea leading to the following theorem. Again, we consider any simulation
protocol for the target correlation $T$ executed by 5 separated players
on a finite amount of copies of a resource system from $\mathcal{R}$
that fulfills the input dependence constraint mentioned before.

\begin{thm}\label{there is a constant factor classical protocol}
  There exists a constant $c$ such that from any simulation protocol
  with distance $\delta$ the existence of a local system with distance
  at most $c\delta$ follows.
\end{thm}

\begin{proof}
  The initial protocol simulates a system $S$ which approximates $T$
  with distance $\delta(S,T)$.

  Consider player 1 first. We build a new protocol by changing player
  1's local output function as follows: Conditioned on input $x_1=0$ and
  any assignment $\rho=\rho(\lambda,b_{n_2+1},\hdots,b_n)$, the new
  function shall constantly output the majority of the outputs of the
  original function under the same condition, that is,
  \begin{align*}
    \hat{f}^1|_\rho&=f^1|_\rho,\\
    \hat{f}^0|_\rho&=\begin{cases}
      1,\text{ if }P(f^0|_\rho=1)\geq \frac{1}{2},\\
      0,\text{ otherwise}.
    \end{cases}
  \end{align*}
  Obviously, $\hat{f}^0$ is independent of the values
  $b_1,\hdots,b_{n_2}$ -- the actual system outputs correlated with
  player 2 -- for it depends only on the related output distribution. One can
  calculate the increase in the simulation distance that is implied by
  this change. Assume a choice of inputs with $x_1=0$ that demands a
  certain correlation with player 1 involved. Conditioned on any $\rho$,
  replacing $f^0$ by $\hat{f}^0$ can increase the chance of violating
  this correlation by at most the probability that the minority output
  is generated by the original function, hence
  $\mu(f^0|_\rho)$. Otherwise, player 1's output behavior has not
  changed at all. So, summing over all possible $\rho$, the change
  increases the simulation distance by at most
  \[
  \sum_{\rho}P(\rho)\mu(f^0|_\rho).
  \]
  Lemma \ref{minimal simulation distance on input 0}, with parameters
  $i=1,j=2$, implies that the distance $\delta(S,T)$ of the original
  simulation protocol is already at least as high. Therefore, our change
  doubles the simulation distance in the worst case. Thus, the new
  simulation protocol approximates $T$ with a distance within
  $2\delta(S,T)$.

  With the same argument on different pairs involving player 1 we change
  the protocol another three times. We sequentially free player 1's
  output function on input $x_1=0$ from the dependence on outputs of
  shared resource systems. Doing so, we obtain a new simulation protocol
  for $T$ with a simulation distance bounded by $2^4\delta(S,T)$. Then,
  we extend this procedure to all players. Finally, we reach a protocol
  where each players' output is independent of the outputs of its shared
  resource systems if getting input 0. The new simulation distance is
  limited by $2^{20}\delta(S,T)$.

  We continue the reasoning with the new simulation protocol. Now assume
  player 1 gets input $x_1=1$. If $x_2=x_5=0$, then correlation
  (\ref{cond5}) of Example \ref{hard target system}, that is,
  \begin{align}
    a_5+a_1+a_2=0~(\text{mod }2),\label{second correlation condition}
  \end{align}
  needs to be fulfilled by the outputs of the players. In the current
  version of the protocol, all players base their outcome solely on the
  shared value $\lambda$ when getting the input $0$, as do players $2$
  and $5$. Conditioned on a $\lambda$, player 1's output function $f^1$
  will, therefore, determine an output not satisfying (\ref{second
    correlation condition}) with a probability of at least
  $\mu(f^1|_\lambda)$. Therefore, the simulation distance of the new
  protocol is bounded from below by the convex combination of these
  values, that is, by
  \begin{align}
    \sum_{\lambda}P_{\Lambda}(\lambda)\mu(f^1|_\lambda).\label{maximal
      distance in the updated protocol}
  \end{align}
  For each $\lambda$, $f^1$ is now replaced by a function evaluating to
  the majority decision of $f^1|_{\lambda}$, similarly to the
  modifications made earlier. This change establishes that player 1's
  output function is independent of the outputs of all shared resource
  systems $R_1,\hdots,R_n$. For each $\lambda$, which is the shared
  value with chance $P_{\Lambda}(\lambda)$, this causes a simulation
  distance increase which is maximally as large as the probability of
  player 1's minority decision given $x_1=1$ and $\lambda$, that is,
  $\mu(f^1|_\lambda)$. The total increase is then maximally as large as
  (\ref{maximal distance in the updated protocol}). Therefore, the new
  protocol, in which player 1's strategy relies only on shared
  randomness, is at most twice as faulty as the old one.

  With the same argument we handle players' $2,3,4$ and $5$'s dependence
  on resource system outputs. As seen above we pay these changes with a
  factor of at most $2^4$ in the simulation distance. Therefore, the
  final protocol for $T$ relies only on shared randomness and simulates
  a certain local system $S'$ with $\delta(S',T)\leq
  2^{25}\delta(S,T)$.\end{proof}

\begin{cor}\label{no bipartite unit for non-adaptive simulations}
  There is at least one multipartite system that cannot be approximated
  arbitrarily well with a non-adaptive protocol on bipartite systems.
\end{cor}

\begin{proof}
  An example is $T$. Assume that there is a non-adaptive protocol on
  resources from $\mathcal{P}^b$ simulating a system $S$ that
  approximates $T$ arbitrarily well, that is, with any
  $\delta(S,T)>0$. We replace all bipartite systems used in this
  protocol by simulations on the resource set $\mathcal{R}$. This can be
  achieved asymptotically perfect by a combination of simulation
  protocols demonstrated in Section \ref{bip unit}. It is a fact that
  one of the needed reductions ($\mathcal{R}\leadsto^\e\mathcal{D}$)
  uses adaptive protocols. However, since these simulations are
  bipartite, the non-adaptive nature of the assumed protocol on
  $\mathcal{P}^b$ translates into the constrained input dependence as
  required in Lemma \ref{minimal simulation distance on input 0} and
  Theorem \ref{there is a constant factor classical protocol}.

  It then follows directly from Theorem \ref{there is a constant factor
    classical protocol} that there is a local system that approximates
  $T$ to an arbitrary precision. This contradicts the established fact
  that $T$ violates a Bell inequality~\cite{DiVincenzo-97}.
\end{proof}

\subsection{Adaptive protocols}

Next, we consider the general case where the players of the simulation
game are allowed to agree on any kind of protocol. We will show that the
same example, which was impossible to approximate in the restricted
setting, is a hard instance in the general case as well. This will be
demonstrated by deriving a lower bound on the distance that occurs when
five players simulate the target $T$ on shared randomness and a finite
amount of copies of systems from $\mathcal{R}$. We find a distance bound
that can theoretically reach any small value but decreases rather
slowly, that is, reciprocally in the number of shared resources.

As a first step into this direction, we show a weaker lower bound that
has not yet the required property but will be useful later. For
$i\in[n]$, let $\delta(S,T|i)$ denote the simulation distance of a
protocol conditioned on the assumption that any player's variable $b_i$
is determined locally and uniformly at random.

\begin{lem}\label{minimal simulation error on random bit}
  Suppose any choices $j\in[5]$, $x_j\in\{0,1\}$ and $i\in[n]$. In a
  simulation protocol for $(T,\mathcal{R})$ in which player $j$'s binary
  variable $b_i$ is assigned the result of a local, fair coin toss, we
  have that
  \begin{align*}
    \delta(S,T|i)\geq\sum_{\rho\in
      S_i(x_j)}P(\rho)[\mu(f^{x_j}|_\rho)&-\tfrac{1}{2}\mu(f^{x_j}|_{\rho,b_i=0})\\
    &-\tfrac{1}{2}\mu(f^{x_j}|_{\rho,b_i=1})].
  \end{align*}
\end{lem}

\begin{proof}
  Informally spoken, the conditional distance $\delta(S,T|i)$ is at
  least half as high as the minimal probability, for any fixed inputs,
  that player $j$'s output changes depending on his local bit $b_i$,
  while the outputs of the other players remain constant.

  For the rest of the proof we fix the players' inputs such that their
  outputs have to satisfy a correlation from (\ref{cond1}) --
  (\ref{cond6}) with player $j$ involved. Let $\rho_j$ denote an
  assignment of values to all outputs of resource systems that are
  received by players $[5]\backslash\{j\}$. Fixing $\rho_j$ and the
  shared random value $\lambda$ determines the output of these
  players. If player $j$'s output remains variable, then a violation of
  the required correlation occurs. Given $x_j$ and an assignment
  $\rho\in S_i(x_j)$ (see Definition \ref{definition of the assignment
    set}), it is convenient to introduce the random variables
  $\xi=f^{x_j}|_{\rho,b_i=0}$ and $\zeta=f^{x_j}|_{\rho,b_i=1}$. We have
  $P(\xi=a)=P(f^{x_j}=a|b_i=0,\rho)$ and
  $P(\zeta=a)=P(f^{x_j}=a|b_i=1,\rho)$ forn all outputs
  $a\in\{0,1\}$. The probability of the described violation can then be
  stated as
  \begin{align}
    \delta(S,T|i)\geq\frac{1}{2}\sum_{\rho\in
      S_i(x_j)}\sum_{\rho_j}P(\rho,\rho_j)P(\xi\neq \zeta|\rho_j).
    \label{original expression for error}
  \end{align}
  Here, $P(\xi\neq \zeta|\rho_j)$ denotes the probability that player
  $j$'s output when obtaining $\rho$ and then $b_i=0$ differs from the
  output in the case $\rho$ and $b_i=1$, conditioned on $\rho_j$. Assume
  $\rho$ to be fixed for the moment. Since $f^{x_j}$ is a Boolean
  function, $P(\xi\neq \zeta|\rho_j)$ can be decomposed into two basic
  cases.
  \begin{align}
    P(\xi\neq\zeta|\rho_j)=P(\xi=0,\zeta=1|\rho_j)+P(\xi=1,\zeta=0|\rho_j).
    \label{decomposed error formula}
  \end{align}
  Next, we observe that for all $a$ the events $\xi=a$ and
  $\zeta=\bar{a}$ conditioned on $\rho_j$ are independent because
  $b_i=0$ and $b_i=1$ are mutually exclusive conditions. Therefore,
  \begin{align}
    P(\xi=a,\zeta=\bar{a}|\rho_j)=P(\xi=a|\rho_j)P(\zeta=\bar{a}|\rho_j)\label{expression
      xy}
  \end{align}
  holds for any $\rho_j$. From now on let $a$ be such that
  $\mu(f^{x_j}|_\rho)=P(f^{x_j}=a|\rho)$, that is, $a$ is the least
  probable output under the condition $\rho$. It is easy to see that
  (\ref{expression xy}) can be rewritten to
  \begin{align*}
    P(\xi=a,\zeta=\bar{a}|\rho_j)=&P(\xi=a|\rho_j)\\
    &-P(\xi=a|\rho_j)P(\zeta=a|\rho_j)
  \end{align*}
  and therefore, applied to (\ref{decomposed error formula}), we get the
  equality
  \begin{align*}
    P(\xi\neq\zeta|\rho_j)&=P(\xi=a|\rho_j)+P(\zeta=a|\rho_j)\\
    &-2P(\xi=a|\rho_j)P(\zeta=a|\rho_j)\\
    &=2P(f^{x_j}=a|\rho,\rho_j)\\
    &-2P(\xi=a|\rho_j)P(\zeta=a|\rho_j).
  \end{align*}
  Now we choose the bit $b_i$ such that the term
  $\sum_{\rho_j}P(\rho_j|\rho)P(f^{x_j}=a|\rho,b_i,\rho_j)$ is minimal,
  one easily obtains the lower bound
  \begin{align*}
    P(\xi\neq\zeta|\rho_j)\geq
    2P(f^{x_j}=a|\rho,\rho_j)-2P(f^{x_j}=a|\rho,b_i,\rho_j)
  \end{align*}
  for any $\rho_j$. Let $\bar{\rho}$ denote any assignment of the
  remaining variables in $b_1,\hdots,b_n$ not fixed by $\rho$. Observe
  that for any assignments $\rho,\rho_j$ we have
  \begin{align*}
    P(f^{x_j}=a|\rho,\rho_j)&=\sum_{\bar{\rho}}P(\bar{\rho},f^{x_j}=a|\rho,\rho_j)\\
    &=\sum_{\bar{\rho}}P(\bar{\rho}|\rho,\rho_j)\mathbf{1}_{f^{x_j}=a}(\rho,\bar{\rho}),
  \end{align*}
  where $\mathbf{1}_{f^{x_j}=a}(\rho,\bar{\rho})\in\{0,1\}$ indicates
  whether $f^{x_j}(\rho,\bar{\rho})$ evaluates to $a$. For any fixed
  $\rho$ this implies
  \begin{align*}
    \sum_{\rho_j}&P(\rho_j|\rho)P(f^{x_j}=a|\rho,\rho_j)\\
    &=\sum_{\rho_j}P(\rho_j|\rho)\sum_{\bar{\rho}}P(\bar{\rho}|\rho,\rho_j)\mathbf{1}_{f^{x_j}=a}(\rho,\bar{\rho})\\
    &=\sum_{\rho_j,\bar{\rho}}P(\rho_j,\bar{\rho}|\rho)\mathbf{1}_{f^{x_j}=a}(\rho,\bar{\rho})\\
    &=\sum_{\bar{\rho}}P(\bar{\rho}|\rho)\mathbf{1}_{f^{x_j}=a}(\rho,\bar{\rho})=P(f^{x_j}=a|\rho).
  \end{align*}
  Thus, (\ref{original expression for error}) becomes
  \begin{align*}
    \delta(S,T|i)\geq&~
    \sum_{\rho}P(\rho)P(f^{x_j}=a|\rho)\\
    &-\sum_{\rho,\rho_j}P(\rho,\rho_j)P(f^{x_j}=a|\rho,b_i,\rho_j).
  \end{align*}
  This is almost the representation we seek. Now, as a last step, we
  will get rid of the second dependence on $\rho_j$ by using the initial
  assumption on $b_i$. Let now $\bar{\rho}$ stand for any assignment of
  the remaining variables in $b_1,\hdots,b_n$ not fixed by $\rho$ and
  not equal to $b_i$. Observe that for any assignments $\rho,\rho_j$ and
  any $b_i$ it holds that
  \begin{align*}
    P(f^{x_j}=a|\rho,b_i,\rho_j)&=\sum_{\bar{\rho}}P(\bar{\rho},f^{x_j}=a|\rho,b_i,\rho_j)\\
    &=\sum_{\bar{\rho}}P(\bar{\rho}|\rho,b_i,\rho_j)\mathbf{1}_{f^{x_j}=a}(\rho,b_i,\bar{\rho}).
  \end{align*}
  Using the equality $P(\rho_j|\rho)=P(\rho_j|\rho,b_i)$ -- which holds
  only because $b_i$ is assumed to be determined locally at random,
  meaning that $P(b_i)=P(b_i|\rho)=P(b_i|\rho,\rho_j)$ -- yields
  \begin{align*}
    \sum_{\rho_j}&P(\rho_j|\rho)P(f^{x_j}=a|\rho,b_i,\rho_j)\\
    &=\sum_{\rho_j,\bar{\rho}}P(\rho_j,\bar{\rho}|\rho,b_i)\mathbf{1}_{f^{x_j}=a}(\rho,b_i,\bar{\rho})\\
    &=\sum_{\bar{\rho}}P(\bar{\rho}|\rho,b_i)\mathbf{1}_{f^{x_j}=a}(\rho,b_i,\bar{\rho})=P(f^{x_j}=a|b_i,\rho).
  \end{align*}
  Therefore, we can reformulate (\ref{original expression for error}) to
  \begin{align*}
    \delta(S,T|i)\geq\sum_{\rho}P(\rho)[P(f^{x_j}=a|\rho)-P(f^{x_j}=a|b_i,\rho)].
  \end{align*}
  Remember that the variable $b_i$ has been assigned such that for each
  $\rho$ the probability $P(f^{x_j}=a|b_i,\rho)$ is smaller or equal to
  $P(f^{x_j}=a|\bar{b}_i,\rho)$. Therefore, using
  $\mu(f^{x_j}|_\rho)=P(f^{x_j}=a|\rho)$ it is easy to derive the
  equality $P(f^{x_j}=a|b_i,\rho)=\mu(f^{x_j}|_{\rho,b_i})$. Also,
  $\mu(f^{x_j}|_{\rho,b_i})\leq\mu(f^{x_j}|_{\rho,\bar{b}_i})$ holds
  since otherwise $\mu(f^{x_j}|_\rho)=P(f^{x_j}=a|\rho)$ would not be
  satisfied. From this one can conclude
  \begin{align*}
    \delta(S,T|i)\geq\sum_{\rho}P(\rho)[\mu(f^{x_j}|_\rho)&-\tfrac{1}{2}\mu(f^{x_j}|_{\rho,b_i=0})\\
    &-\tfrac{1}{2}\mu(f^{x_j}|_{\rho,b_i=1})]
  \end{align*}
  which finishes the proof.\end{proof}

For a fixed $i$, the lower bound on the simulation error as given in
Lemma \ref{minimal simulation error on random bit} is independent of the
number of resource systems used in the simulation and can be trivial,
that is, equal to zero. The argument may be boosted by calculating a
lower bound to the sum of all conditional distances. As we will see now
this approach finally yields the linear dependence on the number of used
resource systems.

Suppose an adaptive simulation protocol for $T$ on resources from
$\mathcal{R}$ in which each player shares at most $n$ resource systems.

\begin{lem}\label{at least one conditional distance is big}
  For any choices of $j\in[5]$ and $x_j\in\{0,1\}$, there is at least
  one index $i\in[n]$ such that
  \[
  \delta(S,T|i)\geq\frac{1}{n}\sum_{\lambda}P_\Lambda(\lambda)2\mu(f^{x_j}|_\lambda).
  \]
\end{lem}

\begin{proof}
  For any player $j$, we will show a lower bound to the sum of
  conditional distances, assuming for each summand a protocol in which
  the variable $b_i$ is determined locally and uniformly at random.

  We will use $f$ for $f^{x_j}$ to avoid unnecessary lengths in the
  formulas. After fixing the inputs for the rest of the players
  accordingly, Lemma \ref{minimal simulation error on random bit}
  implies
  \begin{align*}
    \sum_{i=1}^n\delta(S,T|i)\geq\sum_{i=1}^n\sum_{\rho\in
      S_i(x_j)}P(\rho)[\mu(f|_{\rho})&-\tfrac{1}{2}\mu(f|_{\rho,b_i=0})\\
    &-\tfrac{1}{2}\mu(f|_{\rho,b_i=1})].
  \end{align*}
  Introducing player $j$'s local function $\hat{i}=i_{l(\rho)}$ that,
  based on current local assignments $x_j,\rho$, decides the index of
  the next (the $l(\rho)$th) system to use, we replace the right-hand
  side by
  \begin{align*}
    \sum_{\rho\in
      S(x_j)}P(\rho)[\mu(f|_{\rho})-\tfrac{1}{2}\mu(f|_{\rho,b_{\hat{i}}=0})-\tfrac{1}{2}\mu(f|_{\rho,b_{\hat{i}}=1})],
  \end{align*}
  where the set of all assignments is denoted by
  $S(x_j)=\bigcup_{i=1}^nS_i(x_j)$. Observe now that for any $\rho\in
  S_i(x_j)$ and any assignment to $b_i$, which means for all
  $(\rho,b_i=0),(\rho,b_i=1)\in S(x_j)$, we have
  \[
  P(\rho,b_i)\mu(f|_{\rho,b_i})=\tfrac{1}{2}P(\rho)\mu(f|_{\rho,b_i})
  \]
  since $P(\rho,b_i)=P(\rho)/2$. Furthermore, we will make use of the
  fact that for all $\rho\in S(x_j)$, which fix $n-1$ local variables
  from $b_1,\hdots,b_n$ and the shared value $\lambda$, we have
  $\mu(f|_{\rho,b_i=0})=\mu(f|_{\rho,b_i=1}))=0$. Therefore, all
  summands cancel each other out except the ones corresponding to
  initial assignments that fix only $\lambda$. Thus,
  \begin{align*}
    \sum_{i=1}^n\delta(S,T|i)&\geq
    \sum_{\lambda}P_{\Lambda}(\lambda)\mu(f|_{\lambda})
  \end{align*}
  which completes the proof.\end{proof}

It seems natural that the sum
$\sum_{\lambda}P_\Lambda(\lambda)2\mu(f^{x_j}|_\lambda)$, for at least
one player $j\in[5]$, cannot be smaller than some constant related to
the distance between $T$ and the closest local system. Otherwise, this
local system would conflict with the fact that $T$ violates a Bell
inequality. This intuition is formulated and confirmed by the following
theorem.

\begin{thm}\label{slow distance decrease in the general case}
  Any simulation protocol for $T$ in which each player shares maximally
  $n$ resources from $\mathcal{R}$ with the rest of the players
  simulates a system $S$ with
  \[
  \delta(S,T)\in\Omega(n^{-1}).
  \]
\end{thm}

\begin{proof}
  We take the perspective of player $j\in[5]$, such that the sum
  \[
  \sum_{\lambda}P_\Lambda(\lambda)\mu(f^{x_j=0}|_\lambda)
  \]
  is maximal. Player $j$ shares $n_j\leq n$ systems with the rest of the
  players. According to Lemma \ref{at least one conditional distance is
    big}, there exists an index $i\in[n_j]$ identifying one of player
  $j$'s local variables, such that
  \[
  \delta(S,T|i)\geq
  \frac{1}{n}\sum_{\lambda}P_\Lambda(\lambda)2\mu(f^0|_\lambda).
  \]
  The system $R_i$ is shared with another player. It follows directly
  from the $T$'s correlation conditions (Example~\ref{hard target
    system}), that for $x_j=0$ it is always possible to choose inputs to
  the rest of the players such that player $j$ is, and the player with
  which system $R_i$ is shared is not, involved in the required
  correlation. By the nonsignaling principle we can assume that the
  other player with access to $R_i$ completely ignores its end of the
  system while sustaining the original simulation of the required
  correlation. Therefore, player $j$'s variable $b_i$, the output of
  $R_i$, can be interpreted as a local bit distributed uniformly at
  random. Thus, we obtain a lower bound to the simulation distance:
  \begin{align}
    \delta(S,T)\geq\delta(S,T|i).
    \label{distance bound}
  \end{align}
  As in the non-adaptive case we change the local output functions of
  all players to depend only on shared randomness $\lambda$, when given
  the input $0$. Now, each player decides on the outcome with the
  greater probability, when conditioning its corresponding output
  function on $\lambda$. Thus, for each value of $\lambda$, a departure
  from the original protocol is as probable as a change in a player's
  outcome behavior -- the probability of the minority decision under
  $\lambda$. It follows that they now simulate a system $S'$ with
  \[
  \delta(S',T)\leq\delta(S,T)+5\sum_{\lambda}P_\Lambda(\lambda)\mu(f^0|_\lambda).
  \]
  Suppose this change has been adapted and the new simulation protocol
  is now executed instead. In a situation where player $j'$ is given
  $x_{j'}=1$, and the two other players also involved in the required
  correlation receive the input 0 (there is such a correlation
  condition for each choice $j'\in[5]$), the correct output of player
  $j'$ is determined as soon as $\lambda$ is fixed. Therefore,
  \[
  \delta(S',T)\geq\max_{j'\in[5]}\sum_{\lambda}P_\Lambda(\lambda)\mu(f^{x_{j'}=1}|_\lambda).
  \]
  If we change the players' output functions again, in such a way that
  they depend only on shared randomness given input $1$ as well, then
  they simulate a purely classical (local) system $S_c$ with a
  simulation distance of
  \begin{align*}
    \delta(S_c,T)&\leq\delta(S,T)+5\sum_{\lambda}P_\Lambda(\lambda)\mu(f^0|_\lambda)\\
    &\hspace{1.55cm}+5\max_{j'\in[5]}\sum_{\lambda}P_\Lambda(\lambda)\mu(f^{x_{j'}=1}|_\lambda)
    \\
    &\leq\delta(S,T)+5\sum_{\lambda}P_\Lambda(\lambda)\mu(f^0|_\lambda)+5\delta(S',T)
    \\
    &\leq
    6\delta(S,T)+30\sum_{\lambda}P_\Lambda(\lambda)\mu(f^0|_\lambda),
  \end{align*}
  which is at least as high as the minimal distance $\delta_c$ that can
  be achieved with a classical protocol. We conclude that either the
  original simulation protocol is bounded away from 0, for example by
  $\delta(S,T)>\delta_c/12$, or otherwise, if
  $\delta(S,T)\leq\delta_c/12$, then we have
  \[
  \frac{\delta_c}{2}\leq
  30\sum_{\lambda}P_\Lambda(\lambda)\mu(f^0|_\lambda).
  \]
  Using inequality (\ref{distance bound}), this yields that any
  simulation protocol for $T$ on resources from $\mathcal{R}$ has a
  distance from the specification of
  \begin{align*}
    \delta(S,T)&\geq
    \frac{1}{n}\sum_{\lambda}P_\Lambda(\lambda)\mu(f^0|_\lambda)\geq\frac{1}{n}\cdot\frac{\delta_c}{60}.
  \end{align*}
\end{proof}

Theorem \ref{slow distance decrease in the general case} implies that
the error in any simulation of $T$ maximally decreases at rate that is
reciprocal to the number of shared resources between two players. Hence,
no simulation with finitely many bipartite resource systems can be
perfect -- a known fact that has already been proved in
\cite{pironio-2005}. Furthermore, it gives evidence that some
multipartite quantum correlations are at least quite costly to
approximate.

\section{Conclusions}\label{concl}

In this work we proved and analyzed a possible way for two separated
parties to transform a supply of shared PR boxes into any desired
bipartite system. The simulation can be made arbitrarily accurate by
increasing the number of resources. This establishes the PR box as a
unit of bipartite nonlocality. An analysis of our scheme's efficiency
reveals that reducing the output alphabet is particularly expensive in
terms of required resource systems. Furthermore, we derive limitations
that any bipartite unit will encounter in the multipartite setting. We
find that the asymptotic simulation of certain quantum correlations is
impossible for players restricted to non-adaptive strategies. In the
general case, we derive a lower bound to the simulation distance that
drops reciprocally in the number of deployed resources. Informally
speaking, this does not prevent adaptive protocols from reaching any
small simulation distance but restricts to trading doubled simulation
quality off at least twice as many resources. Note that
$P_{PR}\in\mathcal{R}$ and $\mathcal{R}\subset\mathcal{P}^b$. Therefore,
the bound also holds for PR boxes as resources. A possible
generalization to arbitrary bipartite resource systems is a task left
for future work. It still remains to identify a nontrivial set of
nonsignaling correlations that can serve as a unit of multipartite
nonlocality in the sense adopted in this work. Our results suggest that
such a system has more than two ends. A promising candidate that comes
to mind is the multipartite version of systems characterized by
permutations -- a generalization of the set $\mathcal{D}$. However, the
simulations derived in this article seem to be unfit for a direct
application.

Simulation protocols that amplify the violation of a CHSH inequality of
given resource systems, so-called nonlocality distillation protocols,
have been introduced in~\cite{fww-2009} and since improved
in~\cite{brunner2009,Hoyer2010,PhysRevA.80.062107}. Here, we want to
point out an interesting implication of these interconversions. The
distillation protocol in~\cite{brunner2009} achieves an asymptotic
simulation of a PR box through processing a finite supply of correlated
nonlocal boxes, which are convex combinations of a PR box and perfectly
correlated random bits. We must, therefore, conclude from Theorem
\ref{main bipartite cor} that each correlated nonlocal box is a unit of
bipartite nonlocality in the same sense as the PR box. One may ask if
inner points of the polytope of binary nonsignaling correlations could
also serve as units. We cannot give a final answer to this
question. However, oppositional evidence exists. If it holds true that
there is no nonlocality distillation protocol for isotropic systems, as
already shown for infinitely many examples in the quantum
region~\cite{dw-2008}, then a negative answer follows directly from the
general upper bound on distillable nonlocality derived in~\cite{f-2011}.

\begin{acknowledgments}
  The authors would like to thank the anonymous referee for comments and
  suggestions. This work was funded by the Swiss National Science
  Foundation (SNSF).
\end{acknowledgments}


\end{document}